\documentstyle[prb,aps,amsfonts,preprint]{revtex}

\begin{document}
\draft
\title{{\em Ab initio} pseudopotential study of Fe, Co, and Ni
employing the spin-polarized LAPW approach}
\author{Jun-Hyung Cho and Matthias Scheffler}
\address{Fritz-Haber-Institut der Max-Planck-Gesellschaft, Faradayweg 4-6,
D-14195 Berlin-Dahlem, Germany}
\date{\today}
\maketitle

\begin{abstract}
The ground-state properties of Fe, Co, and Ni are studied with the
linear-augmented-plane-wave (LAPW) method and norm-conserving pseudopotentials.
The calculated lattice constant, bulk modulus, and magnetic moment
with both the local-spin-density approximation (LSDA) and the generalized
gradient approximation (GGA) are in good agreement with those of
all-electron calculations, respectively.
The GGA results show a substantial improvement over the LSDA results,
i.e., better agreement with experiment.
The accurate treatment of the nonlinear core-valence exchange and correlation
interaction is found to be essential for the determination of the magnetic
properties of $3d$ transition metals.
The present study demonstrates the successful application of the
LAPW pseudopotential approach to the calculation
of ground-state properties of magnetic $3d$ transition metals.
\end{abstract}
\medskip
\narrowtext

\section{INTRODUCTION}
\label{sec:intro}

In the past decade, the pseudopotential method with the
local-density-approximation for the exchange-correlation functional
has had tremendous success in the study
of ground-state properties of nonmagnetic systems.\cite{coh86,pic89}
However, its applications to magnetic systems have been very rare
because of the difficulties in dealing with localized $d$ electrons
within this approach.
This difficulty is due to the rather large overlap of the electron densities
of core and valence electrons, which makes it necessary that the exchange
and correlation interaction of core and valence electrons is taken into
account properly,\cite{bac82,lou82} i.e., a linearization of this term,
which is commonly assumed in pseudopotential studies, is not acceptable.
Indeed, recent studies have shown that partial-core corrected pseudopotentials
provide an accurate description of the ground-state properties of $3d$
transition metals.\cite{zhu92,cho95}

The pioneering pseudopotential calculations of Greenside and
Schl\"{u}ter\cite{gre83}
for ferromagnetic bcc Fe were unfortunately not very successful possibly
due to convergence problems caused by the fact that a Gaussian basis set
was employed: The calculated equilibrium lattice constant and magnetic
moment were
significantly lower than those resulting from all-electron calculations.
Recently, Zhu, Wang, and Louie\cite{zhu92} performed a
mixed-basis pseudopotential calculation for the ground-state properties of
Fe and found good agreement with the results of all-electron calculations;
and Cho and Kang\cite{cho95} found that the ground-state properties
of Ni were described well even when a ``soft'' $d$ pseudopotential is used
together with a plane wave basis set.
These two recent calculations showed the importance of the core-valence
exchange and correlation interaction and that a reliable description of
the structural
and magnetic properties of $3d$ transition metals is indeed provided by the
{\em ab initio} pseudopotential method.

In the present work, we calculate the ground-state properties of Fe, Co, and
Ni using the linear-augmented-plane-wave (LAPW) method and norm-conserving
pseudopotentials.
The purpose of this study is to examine the accuracy of the LAPW
pseudopotential approach for magnetic 3$d$ transition metals in a
systematic way.
Since we employ the LAPW basis which is efficient and accurate to represent
localized as well as extended wave function, we can easily deal with the full
core electron density (used in the nonlinear core-valence exchange and
correlation
interaction) as well as the $d$ valence electron density.
We find that the equilibrium lattice constant, bulk modulus, and magnetic
moment calculated with both the local-spin-density
approximation\cite{hoh64,bar72,cep80} (LSDA) and
the generalized gradient approximation\cite{lan81,per86,per91,per92} (GGA)
are in good agreement with those of all-electron calculations, respectively.
Compared to the LSDA results, the GGA results agree noticeably better with
experiment.
In agreement with previous studies\cite{cho95}
we find that the accurate treatment of the nonlinearity of
the core-valence exchange and correlation interaction is essential for
the proper description of the magnetic moment and magnetic energy.
We show that the LAPW pseudopotential approach is an efficient and highly
accurate method to calculate the ground-state properties of magnetic $3d$
transition metals.
We will compare its advantages and disadvantages to those of the conventional
plane wave method.

The remainder of the paper is organized as follows. In Sec. II, the
calculational
method is described. In Sec. III, we present the calculated ground-state
properties of Fe, Co, and Ni and compare them with previous theoretical
results and experiments. Finally, a summary is given in Sec. IV.

\section{METHOD}
\label{sec:method}

The spin-polarized electronic-structure calculations presented in this work
are performed using the LAPW method\cite{blaha,kohler} and norm-conserving
pseudopotentials\cite{ham89} within both the LSDA and the GGA.
We use the Ceperley-Alder\cite{cep80} (CA) and Perdew-Wang\cite{per91,per92}
(PW91) exchange-correlation functionals for the LSDA and GGA calculations,
respectively.
The nonlocal ionic pseudopotentials of Fe, Co, and Ni are generated from
the ground-state atomic configuration by the generalized norm-conserving
pseudopotential scheme of Hamann.\cite{ham89}
In the present LSDA and GGA calculations we use LSDA pseudopotentials
employing the CA exchange-correlation functional.
For comparison with the GGA results from LSDA pseudopotentials,
we also perform the GGA calculations for Fe using GGA pseudopotentials
with the PW91 exchange-correlation functional.
The nonlinear core-valence exchange and correlation interaction is treated
accurately by using the full core electron density
which is computed in the atomic calculation.
Based on the LAPW method, we expand the wave functions, electron density, and
potential in terms of spherical harmonics inside the muffin-tin (MT)
spheres and
in terms of plane waves in the interstitial region.
The MT sphere radius is chosen to be $R_{\rm MT}$ = 2.1 bohr for all elements
considered in this paper.
For the wave functions we employ spherical harmonics with an angular momentum
up to $l^{\rm wf}_{\rm max}$ = 10 and plane waves up tp a kinetic energy
cutoff of
$\mid {\bf K}^{\rm wf}_{\rm max} \mid^2$ = 13 Ry.
For the electron density and potential we use $l^{\rm pot}_{\rm max}$ = 6 and
$\mid {\bf G}_{\rm max} \mid^2$ = 100 Ry.
The electron density is obtained from the wave functions calculated at 44 and
47 ${\bf k}$ points in the irreducible part of the Brillouin zone for the bcc
and fcc structures, respectively.
These calculational parameters are found to yield a well
converged result for the ground-state properties of Fe, Co, and Ni.

\section{RESULTS}
\label{sec:result}

We start with a discussion of the nonmagnetic (NM) and ferromagnetic (FM)
phases of bcc Fe, fcc Co, and fcc Ni using the CA LSDA functional.\cite{cep80}
The calculated lattice constant, bulk modulus,
and magnetic moment for the FM phase of Fe, Co, and Ni are given
in Table I together with those of previous
calculations\cite{zhu92,leu91,kor92,sin91} and experiments.\cite{kit86}
Note that these other calculations employed different exchange-correlation
functionals.
Our results, when using the LSDA (Ref. 10), show a discrepancy for the
equilibrium
lattice constant, bulk modulus, and magnetic moment of about $-$3 \%, +35 \%,
and
$-$10 \% with respect to the experimental data, respectively.
These errors for 3$d$ transition metals are typical for LSDA calculations.
As shown in Table I, the present LSDA results are in good agreement with
all-electron LSDA calculations such as the
linear-combination-of-atomic-orbitals\cite{leu91}
(LCAO) and linear-muffin-tin-orbital\cite{kor92} (LMTO) methods.
We note that the bulk modulus calculated by our LAPW pseudopotential approach
is systematically smaller than that of other calculations.
We believe that this is possibly due to the use of frozen-core pseudopotentials
and different techniques for solving the one-electron Kohn-Sham equation.

As in previous LSDA calculations,\cite{zhu92,leu91,kor92,sin91} the
present LSDA
study for Fe fails to predict the experimentally observed ground state which
is the FM bcc structure.
Figure 1 shows our results fitted by Murnaghan's equation of state.\cite{mur44}
The NM fcc phase is energetically favored over the FM bcc phase.
The NM bcc phase is found higher in energy than the NM fcc phase by
$\Delta E$ = 24.6 mRy, and the FM bcc phase is higher than the NM fcc phase
by $\Delta E^{\prime}$ = 4.4 mRy.
Comparing these energy differences to other calculations (see Table II)
we find good agreement: The full-potential
linear-augmented-plane-wave\cite{sin91}
(FLAPW) results were $\Delta E$ = 25.8 mRy and
$\Delta E^{\prime}$ = 4.1 mRy, and the mixed-basis pseudopotential
calculations\cite{zhu92} obtained $\Delta E$ = 27 mRy
and $\Delta E^{\prime}$ = 5 mRy.

As it is well known, some ground-state properties of $3d$ transition metals
are not correctly predicted by the LSDA.
In particular for Fe, it has been demonstrated that nonlocal contributions to
the exchange and correlation potential are necessary to predict the correct
ground state.\cite{zhu92,leu91,kor92,sin91}
The previous work employed the PW86 GGA functional.\cite{per86}
In the present work, we perform additional total-energy calculations for
the ground-state properties of Fe, Co, and Ni using the PW91 GGA functional
which fulfills all known sum rules best.\cite{per91,per92}
For these GGA calculations, we use LSDA pseudopotentials calculated
with the CA
exchange-correlation functional.
The fitted energy-volume curves for various phases of Fe are shown in Fig.2
and
the equilibrium lattice constant, bulk modulus, and magnetic moment of Fe, Co,
and Ni are given in the parentheses of Table I.
We find that the GGA provides a substantial improvement over
the LSDA. The increase in the lattice constant and magnetic moment,
and the decrease in the bulk modulus lead to a better agreement with
the experimental values.

In the case of Fe, our GGA calculations predict correctly a FM bcc
ground state (see Fig.2).
We find that the NM bcc phase is higher in energy than the NM fcc phase
by $\Delta E$ = 23.1 mRy, but the FM bcc phase is lower than the NM fcc
phase by
$\Delta E^{\prime}$ = $-$14.9 mRy.
The magnetic energy $E_{\rm mag}$ (i.e., the energy difference between the
NM and
FM phases) is thus 38.0 mRy.
These energetics of Fe compare well with previous
calculations\cite{zhu92,sin91}
(see Table II).
We see that our values are close to those of the all electron
FLAPW\cite{sin91}
calculation ($\Delta E$ = 21.2 mRy, $\Delta E^{\prime}$ = $-$13.9 mRy, and
$E_{\rm mag}$ = 35.1 mRy).
On the other hand, the pseudopotential calculation of Zhu, Wang, and
Louie\cite{zhu92}
gave $\Delta E$ = 23 mRy, and $\Delta E^{\prime}$ = $-$23 mRy.
The latter value is somewhat larger than both the all electron FLAPW and
our present value.
Zhu, Wang, and Louie pointed out that the overestimation of the magnetic
energy
in their GGA calculations was due to the poor transferability of
pseudopotentials within the GGA functional.
In fact, a larger lattice constant with a reduction in the bulk modulus
was found in their GGA results (see Table I).
To check the transferability of GGA pseudopotentials, we calculate
the ground-state properties of Fe using a GGA pseudopotential with
the PW91 exchange-correlation functional, i.e.,
we use the PW91 functional in the atomic Kohn-Sham equation, in the
unscreening of the effective potential of the atom to determine the ionic
pseudopotential, and in the solid-state calculation.
We find that the ground-state properties of Fe change little by using
GGA pseudopotentials. The calculated lattice constant ($a$ = 5.39 bohr),
bulk modulus ($B$ = 1.71 Mbar), and magnetic moment ($M$ = 2.30 $\mu_B$)
are very close to those of our GGA calculations with LSDA pseudopotentials
($a$ = 5.40 bohr, $B$ = 1.69 Mbar, and $M$ = 2.32 $\mu_B$); and the energetics
for the NM bcc, NM fcc, and FM bcc phases (e.g., $\Delta E$ = 23.3 mRy,
$\Delta E^{\prime}$ = $-$14.2 mRy, and $E_{\rm mag}$ = 37.5 mRy) are also
in good agreement with the above LSDA pseudopotential results.
We show that in the case of Fe, both LSDA and GGA pseudopotentials yield
very similar
results when used together with a GGA calculation for the valence electrons.
Thus we speculate that the somewhat larger lattice constant and magnetic
energy
of the mixed-basis pseudopotential calculations\cite{zhu92} may be
attributed to
the basis incompleteness and in particular to an inaccurate treatment of
core-valence overlapping within the mixed basis set.

Finally we investigate the importance of the nonlinearity of the
exchange-correlation
interaction of the core and valence electrons on the ground-state properties
of Fe, Co, and Ni.
The above calculations were done correctly, but we now linearize the
exchange-correlation functional with respect to the interaction between
the core
and the valence electron density. Such a linearization is not necessary,
but it is rather common in pseudopotential calculations.
The calculated lattice constant, bulk modulus, magnetic moment, and
magnetic energy are compared with the correct results in Table III.
The importance of a correct treatment of the exchange-correlation
functional obvious.
In particular, there is a significant difference in the ground-state
properties
of Fe. On the other hand, we note that the lattice constant and bulk
modulus of Ni
are affected only little, but the change in the magnetic energy $E_{\rm mag}$
is substantial (see Table III).
Hence we conclude that an accurate treatment of the nonlinear core-valence
exchange
and correlation interaction is essential for a trustworthy description of
structural,
elastic, and magnetic properties of 3$d$ transition metals.

\section{SUMMARY}
\label{sec:summary}

We studied the ground-state properties of Fe, Co, and Ni using the
LAPW method and norm-conserving pseudopotentials within both the LSDA and
the GGA.
We found that the equilibrium lattice constant, bulk modulus, and
magnetic moment
are in good agreement with previous all-electron calculations.
Compared to the LSDA results, the GGA results show a better ground-state
properties of Fe, Co, and Ni.
We also find that the accurate treatment of the nonlinear core-valence
exchange and correlation interaction is essential for the determination of
the magnetic properties of $3d$ transition metals.
The present results demonstrate well the reliability of the pseudopotential
approach in describing the ground-state properties of magnetic $3d$
transition metals.

There are several advantages for using the present combination of the LAPW
technique
and pseudopotential approach.
This method provides an efficient and accurate treatment of 3$d$
transition metals, for which the localized $d$ electrons are difficult to
handle in
a conventional plane wave basis set.
For a plane wave calculation a GGA pseudopotential gets rather ``hard'',
but such hardness presents no difficulty in our LAPW approach.
In particular, the use of pseudopotentials allows us to avoid inaccuracies
of the LAPW
basis functions due to approximate orthogonalization between semicore and
valence states,
which is often adopted in the all-electron LAPW method.

\acknowledgements

We thank P. Alippi and G. Vielsack for assistance in the initial stages of
the calculations.
One of us (J.H.C.) would like to acknowledge the financial support from
the Korea Science and Engineering Foundation.

\newpage
\begin{table}
\caption{
Calculated lattice constant ($a$), bulk modulus ($B$), and
magnetic moment ($M$)
for the ferromagnetic phase of Fe, Co, and Ni in comparison with previous
calculations and experiments.
The values are from the LSDA calculation (the GGA results are in the
parentheses).
Abbreviations for the LSDA and the GGA represent the used exchange and
correlation
functionals [CA=Ceperley and Alder (Ref. 10), GL=Gunnarsson and Lundqvist
(Ref. 22), vBH=von Barth and Hedin (Ref. 9), PW86=Perdew and Wang (Ref. 12),
PW91=Perdew (Ref. 14)].
}
\begin{center}
\begin{tabular}{|ll|ccc|}
   &                     & $a$ (bohr) & $B$ (Mbar) & $M$ ($\mu_B$) \\ \hline
Fe & LCAO$^a$            & 5.26(5.44) & 2.64(1.74) & 2.08(2.20)  \\
   & LMTO$^b$            & 5.27(5.46) & 2.66(2.15) & 2.28(2.44)  \\
   & FLAPW$^c$           & 5.22(5.44) & 2.51(1.82) & 2.19(2.13)  \\
   & Pseudopotential$^d$ & 5.29(5.60) & 2.41(1.45) & 2.12(2.35)  \\
   & Present study       & 5.22(5.40) & 2.26(1.69) & 2.01(2.32)  \\
   & Expt.$^e$           & 5.42       & 1.68       & 2.22         \\  \hline
Co & LCAO$^a$            & 6.50(6.73) & 2.68(2.14) & 1.50(1.63)  \\
   & LMTO$^b$            & 6.54(6.70) & 2.55(2.44) & 1.62(1.68)  \\
   & Present study       & 6.51(6.69) & 2.37(2.04) & 1.49(1.66)  \\
   & Expt.$^e$           & 6.70       & 1.91       & 1.72        \\  \hline
Ni & LCAO$^a$            & 6.47(6.73) & 2.50(2.08) & 0.59(0.65)  \\
   & LMTO$^b$            & 6.53(6.70) & 2.68(2.53) & 0.62(0.67)  \\
   & Present study       & 6.50(6.68) & 2.39(1.92) & 0.60(0.64)  \\
   & Expt.$^e$           & 6.65       & 1.86       & 0.61        \\
\end{tabular}
\end{center}
$^a$ Reference 18 (LSDA=CA, GGA=PW86). \hspace{0.1cm} $^d$ Reference 5
(LSDA=CA, GGA=PW86). \\
$^b$ Reference 19 (LSDA=GL, GGA=PW91). \hspace{0.1cm} $^e$ Reference 21.  \\
$^c$ Reference 20 (LSDA=vBH, GGA=PW86).
\end{table}

\newpage

\begin{table}
\caption{
Energetics of Fe from LSDA and GGA calculations.
$\Delta E$ denotes the energy difference between the NM bcc and NM fcc phases;
$\Delta E^{\prime}$, between the FM bcc and NM fcc phases.
$E_{\rm mag}$ is the energy difference between the NM and FM phases of bcc Fe.
}
\begin{center}
\begin{tabular}{|ll|rrr|}
                    &            & $\Delta E$ (mRy) & $\Delta E^{\prime}$
(mRy) & $E_{\rm mag}$ (mRy) \\ \hline
FLAPW$^a$           & LSDA (vBH) & 25.8 &    4.1 &  21.7  \\
                    & GGA (PW86) & 21.2 &  -13.9 &  35.1  \\
Pseudopotential$^b$ & LSDA (CA)  & 27   &    5   &  22  \\
                    & GGA (PW86) & 23   &  -23   &  46  \\
Present study       & LSDA (CA)  & 24.6 &    4.4 &  20.2  \\
                    & GGA (PW91) & 23.1 &  -14.9 &  38.0  \\
\end{tabular}
\end{center}
$^a$ Reference 20. \\
$^b$ Reference 5.
\end{table}

\newpage

\begin{table}
\caption{
Importance of non-linearity of the core-valence exchange-correlation functional
on the ground-state properties of Fe, Co, and Ni.
Listed are the lattice constant ($a$), the bulk modulus ($B$), the magnetic
moment ($M$), and the magnetic energy ($E_{\rm mag}$).
Results noted as ``linearized'' are obtained by a linearization of the
exchange-correlation interaction between core and valence electrons.
All tabulated results are from the LSDA calculations.
}
\begin{center}
\begin{tabular}{|ll|cccc|}
   &    & $a$ (bohr)    & $B$ (Mbar)  & $M$ ($\mu_B$) & $E_{\rm mag}$
(mRy) \\ \hline
Fe & linearized xc      & 5.54 & 1.24 & 3.09 & 192.6  \\
   & correct            & 5.22 & 2.26 & 2.01 &  20.2  \\ \hline
Co & linearized xc      & 6.61 & 2.23 & 2.06 &  75.7  \\
   & correct            & 6.51 & 2.37 & 1.49 &   3.2  \\ \hline
Ni & linearized xc      & 6.50 & 2.38 & 0.74 &  14.9  \\
   & correct            & 6.50 & 2.39 & 0.60 &   2.6  \\
\end{tabular}
\end{center}
\end{table}

\begin{figure}
\caption{
Total energy versus volume curves for nonmagnetic (NM) bcc,
ferromagnetic (FM) bcc,
and NM fcc Fe from LSDA calculations.
Energies are relative to the minimum of the NM fcc curve.
}
\label{fig1}
\end{figure}

\begin{figure}
\noindent
\caption{
Total energy versus volume curves for NM bcc, FM bcc, and NM fcc Fe
from GGA calculations.
Energies are relative to the minimum of the NM fcc curve.
}
\label{fig2}
\end{figure}

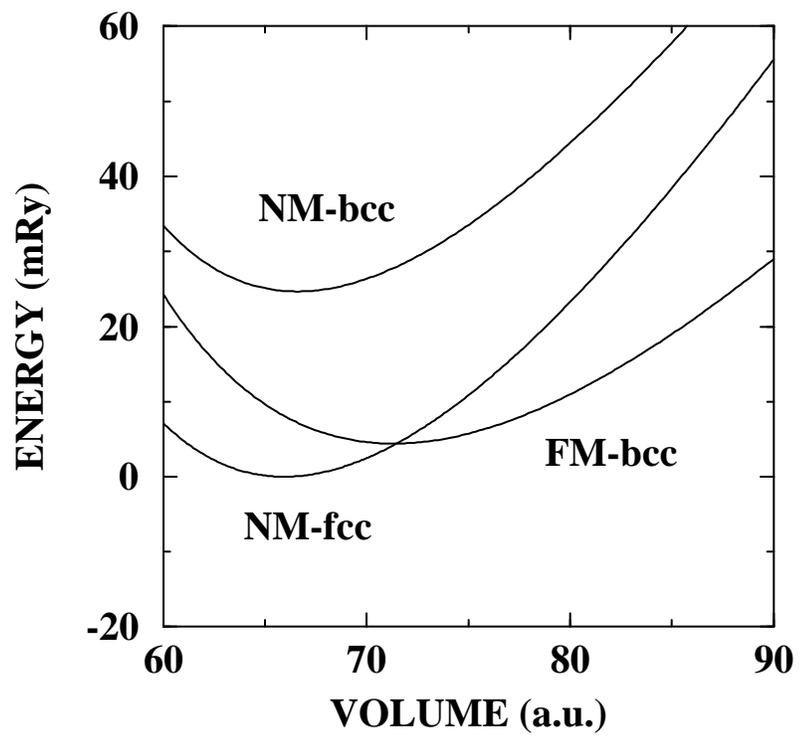
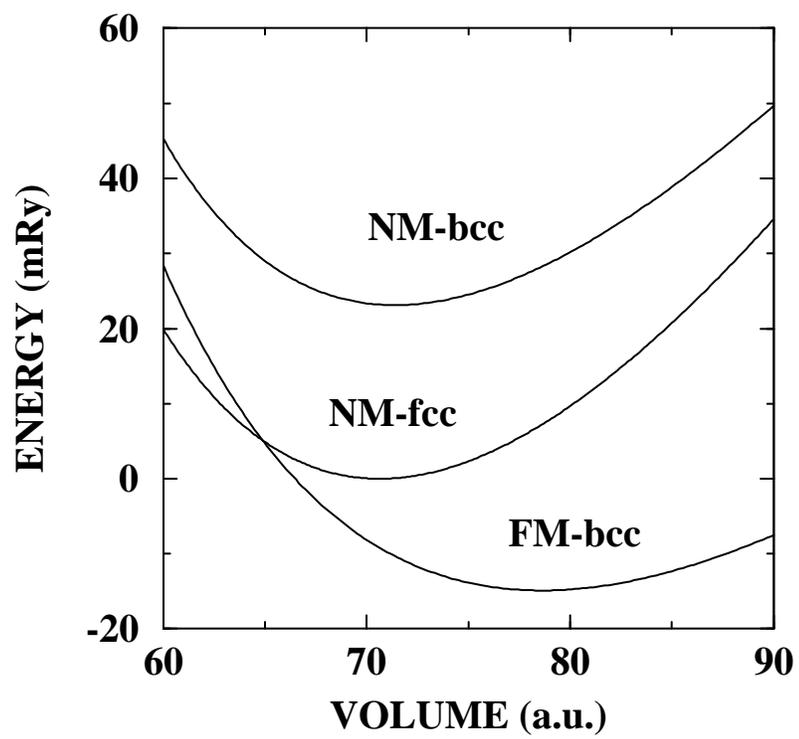
\end{document}